\documentclass[prb, twocolumn, showpacs]{revtex4}

\usepackage{epsfig, amsmath,amssymb}
\input epsf

\begin{document}

\title{Helicoidal Graphene Nanoribbons: Chiraltronics}

\author{Victor Atanasov}
\affiliation{Department of Condensed Matter Physics, \\
Sofia University, 5 blvd. J. Bourchier, 1164
Sofia, Bulgaria}
 \email{vatanaso@phys.uni-sofia.bg}

\author{Avadh Saxena}
\affiliation{Theoretical Division and Center for Nonlinear Studies,
Los Alamos National Laboratory, Los Alamos, NM 87545 USA}
\email{avadh@lanl.gov}

\begin{abstract}
We present a calculation of the effective geometry-induced quantum potential for the 
carriers in graphene shaped as a helicoidal nanoribbon. In this geometry the twist of the 
nanoribbon plays the role of an effective transverse electric field in graphene  and this is 
reminiscent of the Hall effect.  However, this effective electric field has a different sign 
for the two iso-spin states and translates into a mechanism to separate the two chiral  
species on the opposing rims of the nanoribbon.  Iso-spin transitions are expected with 
the emission or absorption of microwave radiation which could be adjusted to be in the 
THz region.
\end{abstract}


\maketitle

{\it Introduction.} The synergy of geometry,  topology and electronic, magnetic or optical properties of 
materials is a prevalent theme in physics, especially when its manifestations are unusual 
and lead to unexpected effects.  Note that helical nanoribbons provide a fertile ground for 
such effects.  Both the helicoid (a minimal surface) and helical nanoribbons are ubiquitous 
in nature;  biomolecules in particular\cite{4, 5, 3, 6}.  A helicoid has two chiralities (Fig. 1).  
Solid state examples include screw dislocations in smectic A liquid crystals \cite{disloc}, 
certain ferroelectric liquid crystals \cite{ferro}, recently synthesized graphene nanoribbons 
\cite{ribbon, ribbon2, ribbon3}, helicoids\cite{helicoid} and spirals\cite{spiral, spiral2}.  
Various physical effects such as electromechanics in graphene nanoribbons and spirals 
including geometric ones can be found in [\onlinecite{ribbonbiblio, ribbon4, ribbon5, ribbon6}].

Novel electronic phenomena in graphene nanoribbons are the main focus here.  In this 
context, our goal is to answer the following question: what kind of effective quantum potential 
do the carriers experience on a graphene helicoid or a helical nanoribbon due to its geometry 
(i.e., curvature and twist)?  Our main finding is that the twist $\omega$ serves as an effective 
electric field acting on the chiral electrons of graphene with a non-vanishing angular 
momentum state.  This is reminiscent of the quantum Hall effect; only here it is geometrically 
induced.  Furthermore, this electric field reverses polarity when the iso-spin (defined below 
with regard to the two components of a Dirac spinor) is changed leading to a separation of 
the iso-spin  states of the carriers on the opposing rims of the nanoribbon. 

The helicoid geometry creates a pseudo-electric field and this unexpected result is intriguing 
in view of the typical effect distortion has on graphene honeycomb lattice, that is to induce a 
pseudo-magnetic field, which leads to the valley-dependent edge states\cite{pseudomagnetic}. 
One possible reason for not observing pseudo-magnetic fields here is that the helicoid is a 
minimal surface (the mean curvature is zero everywhere), that is, it is curved but at the same 
time minimizes the surface energy, therefore not straining the underlying lattice.

We expect our results to lead to new experiments on graphene nanoribbons and other related 
Dirac twisted materials where the predicted effect can be verified and explored in the light of 
spintronics, literally in the case of graphene: ``chiraltronics'' ([\onlinecite{graphenespintronics}] 
and references therein).  

\begin{figure}[t]
\begin{center}
   \includegraphics[scale=0.35]{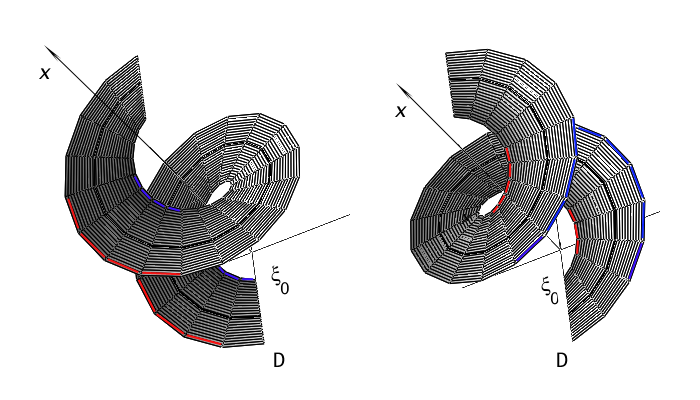}
   \caption{\label{helicoidal} {Two helicoidal nanoribbons with different chiralities: (a) $\omega 
   > 0$ and (b) $\omega < 0$. Vertical axis is along $x$ and the transverse direction $\xi$ is 
   across  the nanoribbons. Here $\xi_0$ is the inner radius and $\rm D$ is the outer radius. 
   The two graphene iso-spin states (color coded as red and blue) collect on opposing rims 
   (separated in space). The respective rims are exchanged when the chirality of the helicoid is 
   reversed. The same exchange takes place when the direction of propagation along the 
   helicoid changes, that is ${\rm m} \to -{\rm m}.$}}
\end{center}
\end{figure}

Note that we treat the nanoribbon as a continuum object without taking into account any 
discreteness of the underlying honeycomb lattice, i.e, we consider a Dirac equation rather 
than a tight-binding model.  Thus are discussion is independent of whether the underlying 
graphene lattice is parallel or perpendicular to the chiral axis, keeping in view the experimental 
observations of Ref. \onlinecite{helicoid}.  We also assume that the helicoid remains a minimal 
surface without any distortion or strain.  Moreover, we assume the stability of the helicoid 
geometry and do not consider any instability issues that may arise experimentally.  

{\it Helicoid geometry.} To elaborate on the geometry of the helicoidal graphene nanoribbon 
we consider a strip whose inner and outer edges follow a helix around the $x$-axis (see Fig. 
\ref{helicoidal} with $\xi_0 =0$).  The represented surface is a {\it helicoid} and is described 
by the following equation:
\begin{equation}
\vec{ r}=x\, \vec{ e}_{x}+ \xi\, [\cos(\omega x)\, \vec{
e}_{y}+ \sin(\omega x)\, \vec{e}_{z}],
\end{equation}
where $\omega = \frac{2\pi {\rm n}}{L}$, $L$ is the total length of the
strip and ${\rm n}$ is the number of $2\pi$-twists. Here
$(\vec{e}_x,\vec{e}_y,\vec{e}_z )$ is the usual
orthonormal triad in $\mathbb{R}^3$ and $\xi\in [0, D]$, where $D$ is the
width of the strip. Let ${ d \vec{r} }$ be the line element and the metric is encoded in
\[
|{d \vec{r}}|^2 = (1 + \omega^2\xi^2)dx^2 + d\xi^2 = h_1^2dx^2 +
h_2^2d\xi^2,
\]
where $h_1=h_1(\xi)=\sqrt{1+\omega^2\xi^2}$ and $h_2=1$ are the Lam\'e
coefficients of the induced metric (from $\mathbb{R}^3$) on the strip. 
Next, we add a comment on the {\it helicoidal 
nanoribbon}, that is a strip defined for $\xi \in [\xi_0, D]$ (see Fig. 
\ref{helicoidal}). All the conclusions still hold true and all of the 
results can be translated using the change of variables
\[
\xi=\xi_0 + s (D-\xi_0), \qquad s\in [0,1].
\]
Here $s$ is a dimensionless variable and one easily sees that for $\xi_0 \to 0$ we again obtain the helicoid.

{\it Effective geometric potential.} In order to answer the question posed above, here 
we study the helicoidal surface to gain a broader understanding of the {\it interaction} 
between Dirac particles and curvature and the resulting possible physical effects. The 
properties of {\it free} electrons on this geometry have been considered before \cite{daCosta, 
Dan*04, us} in the case of Schr\"odinger materials. The results of this paper are based on 
the Dirac equation for a {\it confined} quantum particle on a sub-manifold of $\mathbb{R}^3$.  
Following Refs. [\onlinecite{Jensen, avadh&me1, avadh&me2}] an effective potential 
appears in the two dimensional Dirac equation which in this geometry has the following form:
\begin{eqnarray}\label{2dDirac}
\left(\begin{array}{cc} -k_{+} & \frac{i k_{x}}{\sqrt{1+\omega^2 \xi^2} }  -i \partial_{\xi}    \\  \frac{i k_{x}}{\sqrt{1+\omega^2 \xi^2} }  + i \partial_{\xi}  & -k_{-}\end{array}\right)
\left(\begin{array}{c}\chi_A \\ \chi_B \end{array}\right) &=&0 ,\\
 k_{\pm}= \pm { E }/{\hbar v_F} ,&& 
\end{eqnarray}
where $k_x$ is the partial momentum in $x$-direction. For more information on the derivation, 
refer to the Appendix as well as Ref. [\onlinecite{graphenedirac}].

Let us consider here the azimuthal angle around the $x$ axis: $\omega x$ and the angular momentum
along this axis (cylindrical symmetry):
\begin{equation}
L_x = -  \frac{i \hbar}{\omega} \frac{\partial}{\partial x} .
\end{equation}
This operator has the same eigenfunctions  $L_x \phi(x)=\hbar {\rm m} \phi(x)$
as the Hamiltonian since they commute. The corresponding eigenvalues are $\hbar {\rm m}.$
We conclude that the momentum $k_x$ is quantized
\begin{equation}
k_x= {\rm m} \omega, \qquad {\rm m} \in \mathbb{Z}.
\end{equation}
This is not surprising because of the periodicity of the wave function
along $x$. Note that the value of the angular momentum quantum number
determines the direction the carriers take along the $x$ axis either
upward ${\rm m} > 0$ or downward ${\rm m} < 0.$  This situation is reversed for a
helicoid with opposite chirality (Fig. 1).

\begin{figure}[t]
\begin{center}
   \includegraphics[scale=0.35]{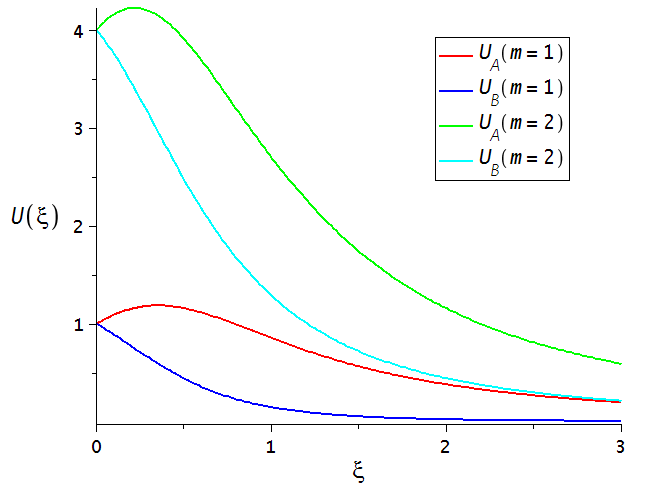}
   \caption{\label{fig: potential} {The potential acting on each of the iso-spin states as a function of the width of the helicoid $\xi$. Here $\omega > 0.$ Note that the potentials have a maximum and then fall off  $\propto 1/\xi^2.$ The extremum for $\rm |m|=1$ state is reached for $\xi_{extr}=1/(\omega\sqrt{8}).$ For $\xi \gg \xi_{extr}$ the iso-spin separation scales as $\Delta U (\xi \gg \xi_{extr}) \approx   \frac{2 {\rm |m|}}{\xi^2}.$}}
\end{center}
\end{figure}

Now we obtain for the first and second components of the spinor, that is the iso-spin states, the following governing effective Schr\"odinger equations: 
 \begin{eqnarray}
-\partial_{\xi}^2 \chi_A + U_A(x) \chi_A &= &- k_{\xi}^2 \chi_A ,\\
-\partial_{\xi}^2 \chi_B + U_B(x) \chi_B &= &- k_{\xi}^2 \chi_B ,\\
k_{\xi}^2 = k_{+} k_{-} &=& - {E^2}/{(\hbar v_F )^2} , 
\end{eqnarray}
where the corresponding potentials are
\begin{eqnarray}
U_A=W_{\rm m}^2  - W'_{\rm m}= \frac{k_{x}^2}{1+\omega^2 \xi^2 } + \frac{k_{x} \omega^2 }{\left(1+\omega^2 \xi^2 \right)^{3/2} } \; \xi ,\\
U_B=W_{\rm m}^2  + W'_{\rm m} = \frac{k_{x}^2}{1+\omega^2 \xi^2 } - \frac{k_{x} \omega^2 }{\left(1+\omega^2 \xi^2 \right)^{3/2} } \; \xi. 
\end{eqnarray}
Here $W_{\rm m}={k_{x}}/\sqrt{1+\omega^2 \xi^2 }.$ These potentials are pseudo-binding and are depicted in Fig. \ref{fig: potential}. Note the qualitative behavior after the extremal point is reached for
\begin{equation}
\xi_{extr}=\frac{1}{| \omega|} \frac{ \sqrt{ 1+ |{\rm m}|^2 - \sqrt{|{\rm m}|^4 -3 |{\rm m}|^2      } } }{\sqrt{2(1-|{\rm m}|^2)}} , 
\end{equation}
provided $|{\rm m}| \neq 1.$ In the case $|{\rm m}|=1$ the extremum is reached for $\xi_{extr}=1/(\omega\sqrt{8}).$

Suppose the width of the nanoribbon $W$ is smaller than $1/(\omega\sqrt{8}),$ that is $W < L/(4 \sqrt{2} \pi {\rm n}),$ then we can  approximate the potential and restrict the expansion to the first order terms
\begin{eqnarray}\label{approx_potentials}
U_A \approx k_{x}^2 + k_{x} \omega^2 \xi, \quad U_B \approx k_{x}^2 - k_{x} \omega^2 \xi , 
\end{eqnarray}
then the governing effective equations become
\begin{eqnarray}
-\partial_{\xi}^2 \chi_A + \left( k^2 +  k_{x} \omega^2 \xi   \right) \chi_A &= & 0 ,\\
-\partial_{\xi}^2 \chi_B + \left( k^2 -  k_{x} \omega^2 \xi   \right) \chi_B &= & 0 ,\\
 k_{x}^2 + k_{\xi}^2 &=& k^2  . 
\end{eqnarray}

Note that the geometry induced potential acting on the two different iso-spin states is similar to the application of a constant electric field $\mathcal{E}$, thus reminiscent of the Hall effect: 
\begin{eqnarray}
U_A \propto  e \mathcal{E}  \xi, \quad U_B \propto  - e \mathcal{E}  \xi,
\end{eqnarray}
where $\mathcal{E}={k_{x} \omega^2}/{e}$, with its sign being different for the different chiral states. Here $e$ is the electron charge. Therefore, {\it $\mathcal{E}$ separates them on the opposing rims of the helicoidal nanoribbon.}
It is exactly this observation that motivates us to assume a mechanism of separation of chiral states in graphene as the basis for a potential new branch of spintronics, namely chiraltronics.

These potentials are a sum of two contributions, an almost constant repulsive part (which pushes the carriers to the outer rim):
$\frac{k_{x}^2}{1+\omega^2 \xi^2 } \approx {\rm m}^2 \omega^2$ and a variable part $ \frac{k_{x} \omega^2 }{\left(1+\omega^2 \xi^2 \right)^{3/2} } \; \xi \approx \omega^{3} {\rm m} \xi$ 
which is repulsive or attractive as a function of the angular momentum quantum number $ {\rm m }$ but more importantly, given ${\rm m} \geq 0$ attractive for iso-spin A (collects on the inner edge) and repulsive for iso-spin B (collects on the outer edge), see (\ref{approx_potentials}).

The action of the first part  $\propto {\rm m}^2 \omega^2$ qualifies it as a centrifugal potential. It pushes a particle to the boundary of the strip. Physically, one may understand the behavior described above using the uncertainty principle: for greater $\xi$ a particle on the strip will have more available space along the corresponding helix and therefore the corresponding momentum (energy) will be smaller than for a particle closer to the central axis.

Since the behavior of the variable part of the potential  $U_B(\xi)$ for a particle with $ {\rm m }\geq 0$ [$U_A(\xi)$ for $ {\rm m}\leq 0$] qualifies 
it as a quantum anti-centrifugal one, it concentrates the corresponding iso-spin carriers around 
the central axis for a helicoid (or the inner rim for a helicoidal nanoribbon). 
Such anti-centrifugal quantum potentials have been  considered for Schr\"odinger materials previously\cite{vic*07}.

We note that the separability of the quantum dynamics along $x$ and 
$\xi$ directions with different potentials points to the existence of 
an effective mass anisotropy for the chiral electrons on the graphene helicoidal surface.

\begin{figure}[t]
\begin{center}
   \includegraphics[scale=0.4]{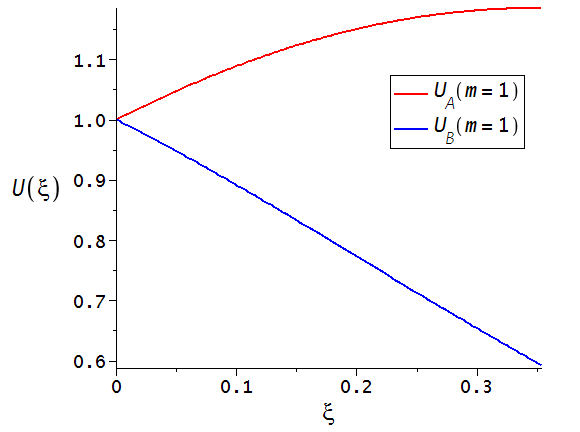}
   \caption{\label{fig:transition region} {Provided the nanoribbon is small enough, so that $\xi < \xi_{extr},$ the potential acting on each of the iso-spin states as a function of the width of the helicoid scales linearly with $\xi.$ Note that the difference between the potentials acting on the two iso-spin states is $\Delta U (\xi < \xi_{extr}) \approx  2 {\rm |m|} |\omega|^3 \xi.$ The frequency of the expected transition is in the THz region (for micron-sized ribbons). See text for further details.}}
\end{center}
\end{figure}

{\it Experimental implications.} A number of experimental consequences can be expected. We begin with the ``thin strip'' case, literally the case in which the width $W < L/(4 \sqrt{2} \pi {\rm n}).$ The pseudo-binding potential (see Fig.~\ref{fig:transition region}) would lead to a two-metastable-states problem and an oscillation between the iso-spin states should be expected. The helicoidal graphene nanoribbon should exhibit an absorption line at frequency $\nu \approx v_F \sqrt{{\rm |m| |n|}^3 2\pi W/L^3}$ connected with the change (positive chirality helicoid $\omega > 0 $) of iso-spin from B to A. Using the restriction on the width of the nanoribbon the frequency turns out to be 
\begin{equation}
\nu \approx  {\rm |n|} \frac{v_F}{L} \; \sqrt{\frac{{\rm |m|}}{2\sqrt{2}} },
\end{equation}
which is determined by the geometric and material properties only. In an attempt to evaluate its order of magnitude we put $L \approx 10^{-6}$ m ($\sim$micron) and $v_F \approx 10^6$ m/s to obtain $\nu \approx 10^{12}$ Hz  well into the THz region. The reverse process is also possible, that is emission in the THz. The change of iso-spin is in this case from A to B. Therefore we might expect a continuous emission, provided we feed the positive chirality helicoid with a current in the inner rim and extract the current (drain it) from the outer rim on the other end. The iso-spin current has to change and therefore emit THz radiation via a standard QED vertex. See the plot of the potential in Fig. \ref{fig:transition region}.

Another experimental effect stems from the form of the geometric potential along the width $\xi$ of the helicoid. The potential in (\ref{2dDirac}) is $
V = {i k_{x}} \sigma_1/ {\sqrt{1+\omega^2 \xi^2} }.$
Here we follow the formalism in Ref. [\onlinecite{Ando}]. The matrix element  of this potential in the Born approximation gives non-vanishing probability $w(\theta)\propto \sin^2({\theta}/{2}),$ where $\theta$ is the scattering angle, for backward scattering. We conclude that the conductivity of the nanoribbon along the width, that is along the rim-to-rim channel is hindered. We believe, this is an additional confirmation of the iso-spin transition the carriers necessarily undertake to populate the opposing rim.

{\it Conclusion.} Our main findings can be summarized as follows: the twist $\omega$ pushes the graphene carriers with iso-spin A and  ${\rm m} \geq 0$ (${\rm m} \leq 0$) towards the outer (inner) edge of the nanoribbon, respectively  iso-spin B  for ${\rm m} \geq 0$ (${\rm m} \leq 0$) towards the inner (outer) edge of the nanoribbon, and {\it effectively separates chiral species on the opposing rims of the helicoid and induces transitions at THz frequencies.}  These results are quite distinct from the ones in the case of twisted Schr\"odinger materials with a scalar wave function and a different geometry induced effective potential\cite{us}. We also predicted an effective mass anisotropy for chiral electrons on the helicoid. We expect our results to motivate new low temperature experiments (in order to restrict to low {\rm m}, that is non-dominant action of the repulsive part of the potential) on twisted graphene nanoribbons in light of the emerging opportunity to separate chiral states, explore chiraltronic applications and possibly create new microwave devices. If the helicoid were elastically deformable then the coupling of chiral electrons to the strain field would possibly lead to a pseudo-magnetic field (in addition to a pseudo-electric field) among other interesting effects. 

In our analysis we have neglected any  effects that may arise due to the underlying  lattice discreteness and distortion (strain) in a real graphene helicoidal nanoribbon.  It would be worthwhile to study these effects numerically along with the potential stability of the considered geometry including the effects of van der Waals adhesion, etc. 

This work was supported in part by the U.S. Department of Energy.

%\section*{Appendix}

{\it Appendix.} The covariant approach for writing the Dirac equation on the curved surface of graphene is the following
\begin{equation}\label{Dirac_curved}
\left( i \hbar v_F \gamma^{\mu} \tilde{\mathcal{D}}_{\mu}  \right)\Psi=0 , 
\end{equation}
where the curvilinear matrices are
\begin{equation}\label{gammas}
\gamma^{\mu} = e_a^{\mu}\tilde{\gamma}^{a}
\end{equation}
and
$ \tilde{\mathcal{D}}_{\mu} = \partial_\mu - \Gamma_\mu.$ Here
\begin{eqnarray}\label{spin_connection_definition}
\Gamma_\mu = \frac14 \; e_{\nu a} \left( \partial_\mu e_{b}^{\nu} + \Gamma_{\mu \lambda} \, ^{\nu} e_{b}^{\lambda}\right)  \; \tilde{\gamma}^{a}\tilde{\gamma}^{b}
\end{eqnarray}
is the spin connection. The Christoffel symbols are defined as:
$\Gamma_{\mu \lambda} \, ^{\nu} = \frac12 \left( \partial_{\mu} g_{\lambda \xi} + \partial_{\lambda} g_{\mu \xi} - \partial_{\xi} g_{\mu \lambda} \right) g^{\xi \nu}.$ The trei-bein fields \cite{Struik} 
\begin{equation}\label{bein_fileds_definition}
g_{\mu \nu }e_a^{\mu} e_b^{\nu}= \eta_{ab}, \qquad 
\eta^{ab }e_a^{\mu} e_b^{\nu}= g^{\mu \nu}
\end{equation}
are defined in terms of the metric on the strip
\begin{eqnarray}\label{metric}
g_{\mu \nu }&=&\left(\begin{array}{ccc}v_F^2 & 0 & 0 \\0 & -(1 + \omega^2\xi^2) & 0 \\0 & 0 & -1\end{array}\right).
\end{eqnarray}
Note, $\eta^{ab }=\eta_{ab}={\rm diag}(1, -1, -1)$ is the choice of the Minkowski metric. Now we define the trei-bein fields $e_a^{\mu}$: 
\begin{eqnarray}
e_a^{t}=\left(\begin{array}{ccc}\frac{1}{v_F} & 0 & 0 \\0 & 0 & 0 \\0 & 0 & 0\end{array}\right),\quad
 e_a^{\xi}=\left(\begin{array}{ccc}0 & 0 & 0 \\0 & 0 & 0 \\0 & 0 & -1\end{array}\right),\\
e_a^{x}=\left(\begin{array}{ccc}0 & 0 & 0 \\0 & -\frac{1}{\sqrt{1 + \omega^2\xi^2} } & 0 \\0 & 0 & 0\end{array}\right)
\end{eqnarray}
and $e_{\mu a}=g_{\mu \nu} e_a^{\nu}.$
The $\gamma^{\mu} = e_a^{\mu}\tilde{\gamma}^{a}$ matrices algebra fulfills $\tilde{\gamma}^{a}\tilde{\gamma}^{b}+\tilde{\gamma}^{b}\tilde{\gamma}^{a}=2 \eta^{ab} \mathbb{I}$ and ${\rm tr}\tilde{\gamma}^{a}=0.$
Upon a straightforward check, the following choice is found to be correct 
\begin{equation}
\tilde{\gamma}^{t}= \sigma_3, \quad \tilde{\gamma}^{x}= i \sigma_1, \quad \tilde{\gamma}^{\xi}= i \sigma_2 ,
\end{equation}
where $\sigma_j$ are the Pauli spin matrices. The curvilinear $\gamma^{\mu}$'s (\ref{gammas}) then are
\begin{equation}
 {\gamma}^{t}=\frac{1}{v_F} \sigma_3, \quad {\gamma}^{x}=-\frac{i \sigma_1}{\sqrt{1 + \omega^2\xi^2} }, \quad {\gamma}^{\xi}= -i\sigma_2 . 
\end{equation}
The non-zero Christoffel symbols components are $\Gamma_{x \xi} \, ^{x} = \Gamma_{\xi x} \, ^{x}  = \frac{\omega^2 \xi}{1+\omega^2 \xi^2}$ and $ \Gamma_{x x} \, ^{\xi}=-\omega^2 \xi.$ As a result, the spin connection $\Gamma_{\mu}$ can be computed from (\ref{spin_connection_definition}) which turns out to be vanishing:
$ \Gamma_{t}=0,$ $\Gamma_{x}=0$ and $\Gamma_{\xi}= 0.$
Putting the corresponding terms in the Dirac equation (\ref{Dirac_curved}) and looking for stationary states with energy $E$,  $\Psi=e^{-\frac{i}{\hbar} E t} \psi$,   we obtain
\begin{equation}
\left(  \frac{\hbar v_F}{\sqrt{1+\omega^2 \xi^2} } \; \sigma_1\partial_{x}+ \hbar v_F \; \sigma_2 \partial_{\xi} \right) \psi=   E \sigma_3 \psi(x, \xi).
\end{equation}
The equations for the iso-spin components after the ansatz  
\begin{equation}
\psi(x, \xi)= \left(\begin{array}{c} \psi_{A} \\ \psi_{B} \end{array}\right),    \; \psi_{A,B}(x, \xi)= e^{i k_{x_1, x_2} x}  \; \chi_{A,B}(\xi)
\end{equation}
are 
\begin{eqnarray}\label{eqns iso}
\left(\begin{array}{cc} k_{+} & i \partial_{\xi}  - i W_{\rm m}(\xi)   \\ -i \partial_{\xi}   - i W_{\rm m}(\xi) & k_{-}\end{array}\right)
\left(\begin{array}{c}\chi_A \\ \chi_B \end{array}\right)=0 , 
\end{eqnarray}
where $ W_{\rm m}(\xi)={k_{x}}/{\sqrt{1+\omega^2 \xi^2} }  $ with the additional condition $ k_{x_1}  =k_{x_2}= k_{x}.$

\end{document}